\documentclass{article}
\usepackage{graphicx}

\begin{document}

\title{Linguistic mechanism of the evolution of amino acid frequencies and genomic GC content}

\author{
\large{Dirson Jian Li} \footnote{Email: {\it dirson@mail.xjtu.edu.cn}} \\
{\it Department of Applied Physics, Xi'an Jiaotong University, Xi'an
710049, PR China}
 }

\maketitle

\abstract{Much information is stored in amino acid composition of
protein and base composition of DNA. We simulated the evolution of
amino acid frequencies and genomic GC content by a linguistic model.
It is showed that the evolution of genetic code determines the
evolution of amino acid frequencies and genomic GC content. We
explained the relationships among amino acid frequencies, genomic GC
content and protein length distribution in a unified theoretical
framework. Especially, the simulations of the evolution of amino
acid frequencies and the codon position GC content agree
dramatically with the results based on the data of all known genomes
so far. Furthermore, we found that the space of average protein
length in proteome and ratio of amino acid frequencies is useful to
describe the phylogeny and evolution. Amazingly, the dots of all the
species in this space form an evolutionary flow. We believe that the
amino acid gain and loss is motivated by the established pattern of
the variation of amino acid frequencies. The linguistic mechanism is
helpful to unveil the origin of the genetic code.}

\newpage
\section*{}

{\Huge A}mino acid frequencies vary slightly among species, while
genomic GC content varies greatly. And proteins from organisms with
GC- (or AT-) rich genomes contain more (or fewer) amino acids
encoded by GC-rich codons \cite{3}\cite{4}. However, the mechanism
of the variation of amino acid frequencies and genomic GC content
was a long-standing and far-reaching problem. We did not realize the
significance of this problem until we explained the profound
relationships among amino acid frequencies, genomic GC content and
protein length distribution in a unified theoretical framework. Its
significance is similar to the role of the cosmic microwave
background in physics, a relic of the evolution of the universe in
early age. The evolution of amino acid frequencies and genomic GC
content decouples from the evolution of the sequences of DNA and
protein. Therefore rich information of the evolution of life in
early time can be stored in the amino acid frequencies and genomic
GC content of extant organisms, which are closely related to the
origin of the genetic code.

Before the life form based on DNA genomes and proteins, there may
exist a simpler life form based primarily on RNA. This earlier era
is referred to as the ``RNA World" when the genetic alphabet of four
bases is thought to have evolved \cite{8}\cite{9}\cite{10}. The
genetic code has been evolving in the context of such a genetic
alphabet while the twenty amino acids joined protein sequences from
the earliest to the latest \cite{11}\cite{12}\cite{13}\cite{14}.
However, in absence of evidence, these problems and all that have to
fall into a twilight zone of speculation and controversy today.
Since no direct data of the evolution of amino acid frequencies and
genomic GC content in early time remains today, we have to choose a
proper evolutionary order for the extant organisms to discern the
regularity of their evolution in early time.

There are different theory about the origin of genetic code
\cite{Jukes}. The frozen-accident theory proposed by Crick states
that any change of the contemporary genetic code would be lethal
unless many simultaneous mutations to alter the code \cite{27}. In
its extreme form, the theory implies that the allocation of codons
to amino acids was entirely a matter of chance. While the
stereochemical theory says that the code is universal because each
amino acid fits its own anticodon or codon in some way. In its
extreme form, the stereochemical theory is said to liken the genetic
code to a ``periodic table'' in which the ``polarity and bulkiness
of amino acid side chains can be used to predict the anticodon with
considerable confidence.'' \cite{periodic_table} As the evolution of
amino acid frequencies is concerned, some believe that it emerged
before the last universal common ancestor of all extant organisms
\cite{5}, while others believe that it conform with the standard,
nearly neutral theoretical expectations \cite{6}. Actually, it is
routinely assumed that amino acid frequencies are constant
\cite{1}\cite{2}. As the difference of genomic GC content among
species is concerned, it is usually explained as the biased AT/GC
pressure exerted on the entire genome during the evolution
\cite{23}\cite{Sueoka 1962}. But the nature of prime biased AT/GC
pressure was unknown. And the mechanism of the correlation of the GC
content between total genomic DNA and the first, second, and third
codon positions was also unknown. Finally as the distribution of
protein length distribution is concerned, what causes the
distribution has not been reported so far.

Considering the analogy between biology and linguistics at the level
of sequence, we proposed a linguistic model to explain all these
phenomena. Storage and expression of the information are crucial in
both biology and linguistics. Like the role of grammar in human
language which can be viewed as a transformation of cell language
\cite{15}, linguistic rules are required to record information in
the protein or DNA sequences. Many attempts have been made to
combine linguistic theory to biology for predicting biological
molecular (RNA, DNA and proteins) structures or trying to
understanding protein assembly and function etc
\cite{7}\cite{16}\cite{17}\cite{18}. Here we focus on the protein
linguistics at early time when genetic code evolved. The closeness
of the result of our linguistic model to the experimental
observations shows that the evolution of genetic code determines the
evolution of amino acid frequencies and genomic GC content. So
linguistics played a significant role in delivering the genetic
information from the RNA world to the DNA-protein world; and the
linguistic mechanism is important for revealing the formation of
genetic code, predicting \cite{7} protein structures and explaining
diversification of life.

\section*{Results}

\subsection*{}

{\bf Evolution of amino acid frequencies.} We analyzed the amino
acid frequencies for 106 species in database with Predictions for
Entire Proteomes (PEP). Although the amino acid frequencies vary
slightly, the pattern of the variation of amino acid frequencies
among extant organisms is well-regulated. According to the consensus
chronology of amino acids to recruit into the genetic code from the
earliest to the latest \cite{19}: G, A, D, V, P, S, E, L, T, R, Q,
I, N, H, K, C, F, Y, M, W, we sort the 106 species by the ratio of
average frequency for 10 later amino acids to average frequency for
10 earlier amino acids (``the ratio of amino acid frequencies'' for
short in the following). Thus, there are 106 data aligning from left
to right in the above order for each of the 20 amino acids. Then we
obtain the evolutionary trends of amino acid frequencies: the
frequencies of G, A, D, V, P, T, R, H, W decrease, while the
frequencies of S, I, N, K, F, Y increase and the frequencies of E,
L, Q, C, M do not vary obviously (Fig. 1a). These variations are
amazingly monotonic by and large. And the magnitude of variation are
different: frequencies of G, A, V, P, R decrease more rapidly than
that of D, T, H, W, while the frequencies of I, N, K, F, Y increase
more rapidly than that of S. Most of the amino acids whose
frequencies decrease (increase) are among the 10 earlier (later)
amino acids, but there are exceptions, i.e., H, W or S. The choice
of order in the above procedure may influence the evolutionary
trends, but the exceptions as well as the magnitudes and
monotonicity of the variation can not be explained by this trivial
reason.

\begin{figure}
\centering{
\includegraphics[width=160mm, height=120mm]{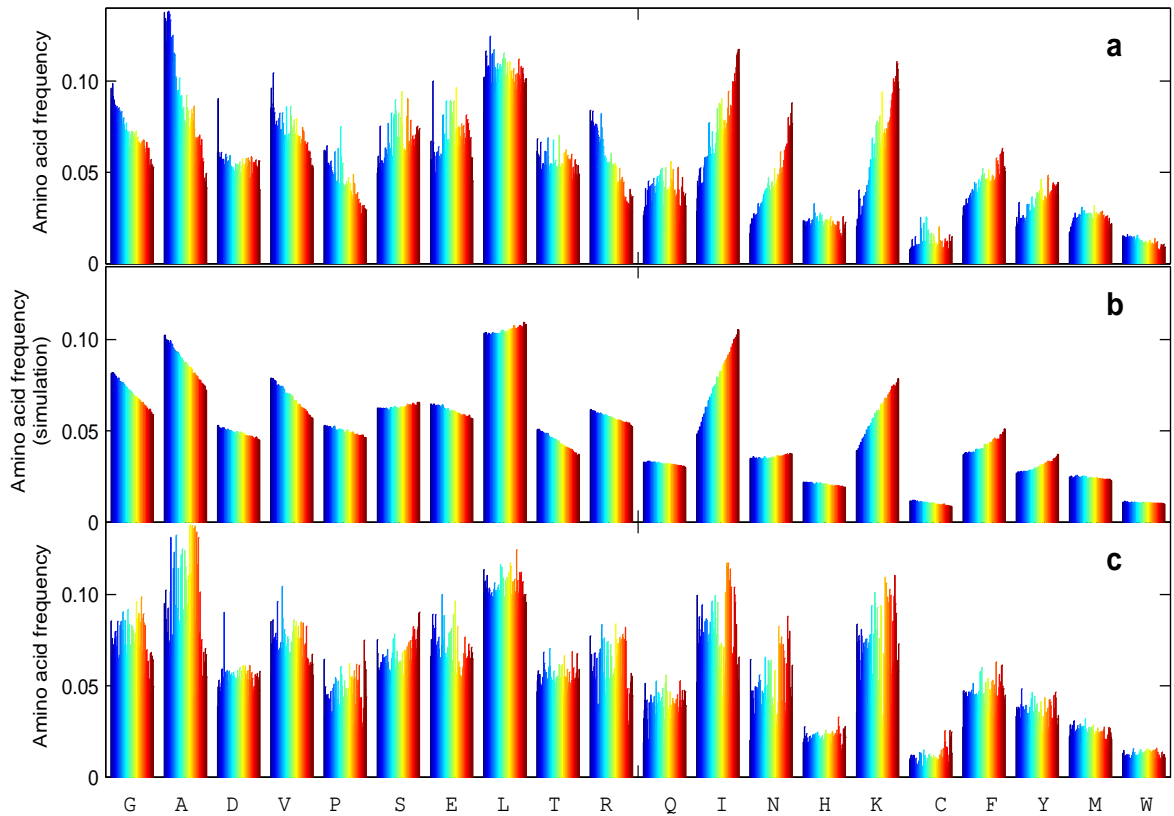}
} \label{fig1} \caption{{\bf Evolution of amino acid frequencies.}
({\bf a}) Evolution of amino acid frequencies based on the data of
106 species, where species are sorted from left to right for each
amino acid by the ratio of average frequency for 10 later amino
acids to average frequency for 10 earlier ones. Amino acids are
aligned chronologically. ({\bf b}) Simulation of the evolution of
amino acid frequencies by the linguistic model, where variant t
increases from left to right for each amino acid. ({\bf c})
Evolution of amino acid frequencies of 106 species sorted by protein
average length.}

\end{figure}

We also obtain the evolutionary trends for each of the three domains
(eubacteria, archaebacteria, and eukaryotes), where the evolutionary
trends for eukaryotes and archaebacteria are decided roughly since
there are only 7 eukaryotes and 12 archaebacteria in PEP. The
evolutionary trends of the three domains are the same for each amino
acid, but the initial amino acid frequencies of the three domains
are different for each amino acid. Take G for example, the initial
amino acid frequencies of eubacteria is greatest while the initial
amino acid frequencies of eukaryote is least. This fact is related
to the phylogeny tree of these three domains, where the deepest
branching separated bacteria from the line leading to archaebacteria
and eukaryote about 3.5 billion years ago and the divergence of
archaebacteria and eukaryote occurred about 2.3 billion years ago
\cite{D}. The frequency of Q for archaebacteria is obviously less
than that for the other two domains.

The results are the same for other orders by the ratio of average
frequency for several later amino acids to average frequency for
several earlier ones. If sorting the species by average protein
length in the proteome (an order independent of the choice of amino
acids), the evolutionary trends are still the same for most of the
amino acids: only the evolutionary trends for P change (Fig.
1c). (It should be noted that a few evolutionary trends are
not obvious in this case because the monotonicity is worse than the
monotonicity when sorting by the ratio of amino acid frequencies.)
The reason is that there is a monotonous relationship (to be
discussed in the following) between protein average length and ratio
of amino acid frequencies, but the deviation of the species from the
midstream of the evolutionary flow are serious (Fig. 4).

We have observed regular variation of amino acid frequencies in the
above. Either the order by ratio of amino acid frequencies or the
order by average protein length reflects an evolutionary direction.
Therefore, it is reasonable to assume that a profound mechanism
underlies the evolution of amino acid frequencies.

\subsection*{}

{\bf The linguistic model.} The evolution of amino acid frequencies
can be explained by our linguistic model, which combines linguistics
and biology substantially. In terms of the tree of genetic code
multiplicity and the genetic code chronology, we propose a model to
simulate the generation of protein and DNA sequences by formal
linguistics. The model consists of three parts: (i) generate protein
sequence by tree adjoining grammar \cite{20}; (ii) set amino acid
for the leaves of grammars in (i) according to the tree of genetic
code multiplicity (see Ref. \cite{21}, which can be obtained by
analyzing the symmetries in the genetic code) with consideration of
the amino acid chronology \cite{19}; and (iii) translate the protein
sequences to the DNA sequences according to genetic code chronology
\cite{22}. (To see materials and methods in detail) The evolution of
genetic code is the core of the model.

There is a variant {\it t} in the model, which determines the
evolution of amino acid frequencies and accordingly represents the
evolution time. A proteome for a species is defined as many a
protein generated by the model with fixed {\it t}, so {\it t} also
identifies species in the model. Thus, the amino acid frequencies
and the average protein length for a species can be calculated. The
evolutionary trends of the amino acid frequencies can be determined
when proteomes are generated at different time {\it t} in the model.
We can also simulate the evolution of genomic GC content after
translating the protein sequences to DNA sequences. We do not
distinguish the three domains in this model.

\subsection*{}

{\bf Simulation of evolution of amino acid frequencies.} The
simulation of our linguistic model (Fig. 1b) coincides with
the global analysis of the data for 106 species (Fig. 1a),
not only in evolutionary trends but in variation magnitudes for most
of the amino acids. The frequencies of G, A, V, R decrease rapidly
while frequencies of D, H, W decrease slowly; and the frequencies of
I, K, F, Y increase rapidly while frequencies of S increase slowly.
The frequencies of E, L, Q, C, M do not vary obviously. Especially,
the simulation for the above exceptions (H, W, S) are good. The
simulations for P, T, N, are not good enough. Such accordance
between theory and biological data can not achieve easily unless we
discover the right mechanism.

Our model depends faithfully on the genetic code multiplicity in
Ref. \cite{21} and the amino acid chronology in Ref. \cite{19}, and
no parameter is added on purpose in the model to alter the trend for
a certain amino acid. Therefore, it is the evolution of genetic
codes (the core of the model) that determines the evolution of amino
acid frequencies. The position of amino acid in the tree of genetic
code multiplicity is crucial to its probability to join the protein
sequence when {\it t} increases, which causes different evolutionary
trends of amino acid frequencies. For example, the positions for G
and A or that for F and Y are equivalent on the tree of genetic code
multiplicity, so the evolutionary trends for G and A or F and Y are
the same.

The 20 amino acid frequencies among the extant organisms are input
as constant parameters. They represent the amount of amino acid in
the primordial soap at the early time when life first appeared. In
simulation, they are just the initial frequencies, when {\it t}
starts, at the leftmost for each amino acid in Fig. 1b. The
evolutionary trends in simulation are not sensitive to the value of
these 20 parameters. When adjusting them to certain extent, the
evolutionary trends do not change. However, when modifying the
genetic code multiplicity a little, the evolutionary trends change
greatly and contradict the trends based on the data of 106 species.
So the genetic code is the key in the linguistic mechanism.

The magnitudes of variation in the simulation are slightly smaller,
on the whole, than the magnitudes based on the data of 106 species
(Fig. 1). So the evolution of amino acid frequencies should
continue after the completion of genetic code, which contributes a
small part to the whole variation.

\subsection*{}

{\bf Co-evolution of amino acid frequencies and GC content and codon
position GC content.} The relationship between genomic GC content
and ratio of amino acid frequencies is regular for the species in
database PEP: genomic GC content decreases linearly with the ratio
of amino acid frequencies (Fig. 2a). The simulation of our model
agrees qualitatively with this relationship (Fig. 2a). In our model,
the evolution of amino acid frequency and the evolution of genomic
GC content are driven by a common variant {\it t}: when {\it t}
increases, there are more later amino acids recruit into the
proteins and the genomic GC content decrease. So the amino acid
frequencies and GC content co-evolved. A protein sequence generated
at later time {\it t} corresponds to the DNA sequence translated
using the later codons, which results in the relationship between
genomic GC content and ratio of amino acid frequencies. The
simulated line does not accord with the optimal line of the dots of
the extant organism because the magnitude of the simulated amino
acid frequency evolution is less than the evolution based on the
data of the extant organism.

\begin{figure}
\centering{
\includegraphics[width=160mm, height=100mm]{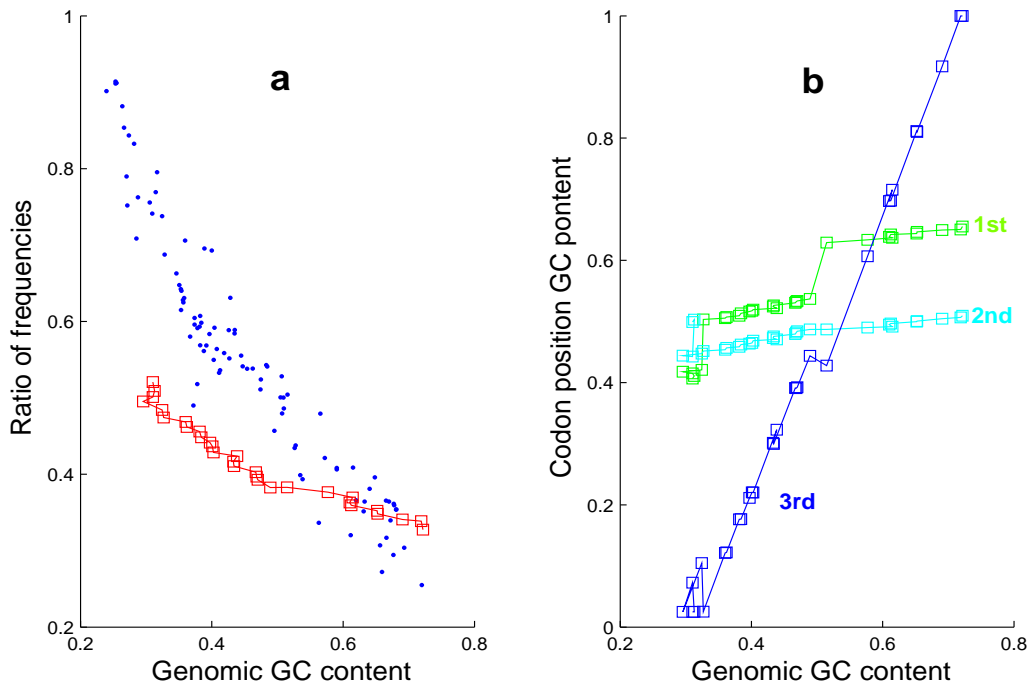}
} \label{fig2} \caption{({\bf a}) Relationship between genomic GC
content and the ratio of average frequency for 10 later amino acids
to average frequency for 10 earlier ones for the species in database
PEP and its simulation by the linguistic model (red solid line).
({\bf b}) Simulation of the correlation of the GC content between
total genomic DNA and the first, second, and third codon positions.}
\end{figure}

The simulation of the correlation of the GC content between total
genomic DNA and the first, second, and third codon positions (Fig.
2b) also agrees with the results based on the data of
organisms in Ref. \cite{23}\cite{GC}, where the correlation slope of
the third codon position is much greater than the correlation slopes
of the first and the second positions and the correlation slopes of
the first position is slightly greater than the correlation slope of
the second position. In the table of codon chronology \cite{22}, G
and C (A and U) occupy all the third positions of earliest (latest)
codons for 20 amino acids, while the bases appear about equally for
the first and second positions. Therefore, the correlation slope for
the first and second positions vary slightly while the slope for the
third position varies greatly. Hence the lower limit of the GC
content (about 1/4) has to equal to one minus its upper limit (about
3/4).

An the simulated results also agree in detail with the results based
on the data of organisms. There is a fine structure, i.e., an

``upward step'', in the middle of the line of the simulated GC
content for the first codon position and a little ``downward step''
in the line of simulated GC content for the third codon position
(Fig. 2b). The predicted fine structure agree dramatically with the
results based on 124 completed eubacterial genomes and 19 completed
archaebacteria genomes \cite{GC}. A convex appears in the line of GC
content for the first codon position and a little concave appears in
the line of GC content for the third codon position (to see fig. 5
in Ref. \cite{GC}). The accordance is more obvious for the results
based on 19 completed archaebacteria genomes. And the lower limits
of the GC content for the first and second codon position are equal
in simulation, which agrees with the results based on data of
organisms. These characters also agree with fig. 2 of Ref. \cite{23}
based on the data of 11 species.

\subsection*{}

{\bf Protein length spectrology.} The linguistic mechanism can also
be supported by the characters of protein length spectrum, or
distribution of protein length. The outline of the spectrum likes a
bell on the whole. But violent fluctuations can be observed all over
the spectrum for each species. When the grammar rules changed in
simulation, the spectrum also changed. Therefore the fluctuation of
the spectrum is not stochastic, which is related to the protein
linguistic rules. Some structures, {\it e.g.} periodic-like
fluctuations \cite{24}, can be observed in the spectra. But we do
not regard it as periodicity. Indeed it is also related to the
underlying grammar rules. A concave appears in protein length
spectrum of each domain at the length near 200 aa \cite{1}, which
might be related to DNA circularization. The main characters of the
spectrum (including bell-shape outline, intrinsic fluctuations,
periodic-like structures) can be simulated by the linguistic model.

\begin{figure}
\centering{
\includegraphics[width=120mm, height=140mm]{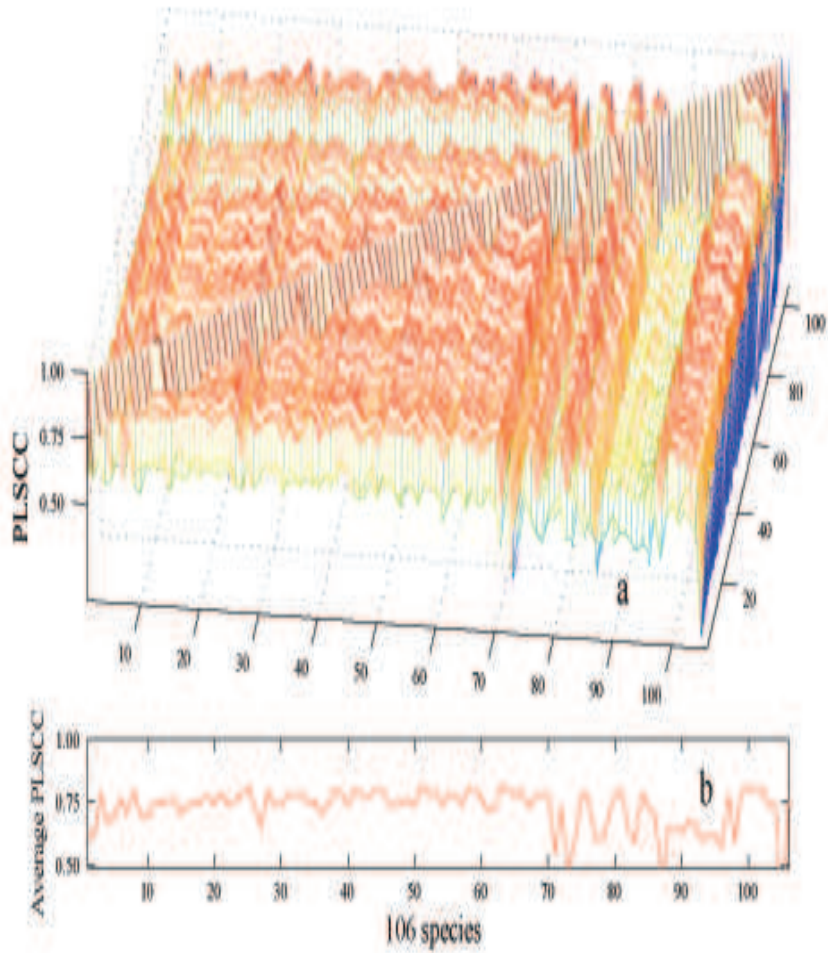}}
\label{fig3} \caption{{\bf Protein length spectrum analysis.} ({\bf a})
PLSCC matrix for 106 species. ({\bf b}) Average PLSCC. }
\end{figure}

In addition, the relationship between species can be inferred by
their protein length spectra. The more closely related pair of
species possesses higher protein length spectrum correlation
coefficient (PLSCC) (defined in materials and methods). The image of
the PLSCC matrix is indeed a global landscape illustrating the
relationship among all species (Fig. 3a), where the
``folded mountains" reflect the branching in the phylogenetic tree.
For example, the number of 71, 89, 94, 95, 97 represent mycoplasma,
which situate at the far end on the phylogeny tree, so the PLSCC for
these species are small (forming a ``valley'' in the landscape).

\subsection*{}

{\bf The evolutionary flow.} We also find the relationship between
the average protein length and the ratio of average frequency for
several later amino acids to average frequency for several earlier
ones. The species of three domains gather together in different
regions in the space of the average protein length and the ratio of
amino acid frequencies (Fig. 4), which supports the three-domain
classification \cite{25}\cite{26}. The distance for closely related
species in this space is small; for instance, the point for human
and the point for mouse in the space are near.

The distribution of all species in the space is a bowed line on the
whole (Fig. 4), which is indeed an evolutionary flow for
the following reasons. The species with large (small) genome or with
big (small) average PLSCC locate in the midstream (margin) of the
flow (Fig. 4); and most of the PLSCC directions (defined in
materials and methods) amazingly parallel with the direction of the
flow by and large (Fig. 4). And the evolutionary direction
of the flow is from the simple organisms toward the complex
organisms. The bowed evolutionary flow can be simulated by our model
(Fig. 4, Embedded), where the time {\it t} gives the right
evolutionary direction of the flow. That the range of ratio of amino
acid frequencies are smaller than the range for 106 species is also
due to the smaller simulated magnitude of evolution of amino acid
frequencies. The bending direction in the simulation agrees with the
flow based on the data of 106 species, which is sensitive to the
exact form of genetic code multiplicity. Such accordance between
theory and data of 106 species confirms the linguistic mechanism of
the protein evolution.

\begin{figure}
\centering{
\includegraphics[width=160mm, height=130mm]{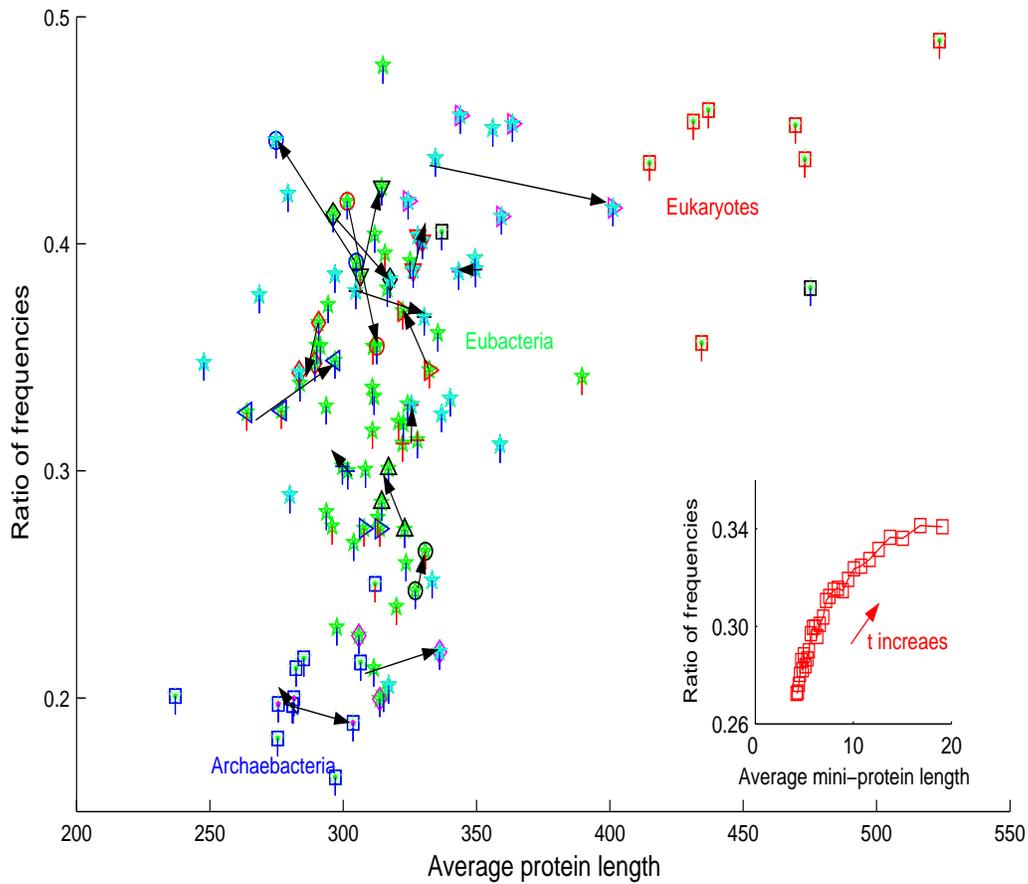}}
\label{fig4} \caption{{\bf The evolutionary flow.} Relationship
between average protein length and ratio of frequencies (here
choosing H, Q, W among later amino acids and G, V among earlier
ones) for 106 species (archaebacteria: blue square, eubacteria:
star, virus: black square, eukaryote: red square). The proteome size
is represented by the colour of tail below each point of species
(big: red, small: blue). The green (caesious) stars denote species
with big (small) average PLSCC. The black arrows denote PLSCC
directions. ({\bf Embedded}) simulation of the bowed evolutionary
flow. }
\end{figure}

In terms of the linguistic model, the variation of amino acid
frequencies and genomic GC content developed mainly in the time when
genetic code evolved. At that time, the three domain had not
separated and most of the extant organisms did not appear, so the
evolutions of amino acid frequencies for the three domains are the
same. Thus, the pattern of the variation of amino acid frequencies
among species must form before the last universal common ancestor of
all extant organisms. Whereafter in the modern evolution, the amino
acid frequencies should continue to evolve motivated by the
established pattern of the difference in amino acid frequencies.
Such a motivation is intrinsic for the evolution of genes. The
evolution of life is a procedure to keep balance between the
restricts by the intrinsic rules of the genes and the demand for
species to adapt the vicissitudes in surroundings. Although the
genes are stored in separate individuals, they are tightly connected
by food chain and sexual network. So the intrinsic rules of the
genes are reflected by the evolutionary flow, and the general
pattern of the variation of amino acid frequencies in early time has
been inherited amongst the extant organisms.

Compared sets of orthologous proteins encoded by triplets of closely
related genomes from 15 taxa, Jordan et al found that C(0.45),
M(0.22), H(0.18), S(0.13), F(0.12) accrue in at least 14 taxa,
whereas P(-0.36), A(-0.16), E(-0.15), Q(-0.10) are consistently
lost, for which they did not give reason (The data in the
parentheses reflect the magnitudes of gain (plus) and loss (minus)
of amino acids) \cite{5}. Their result of the modern evolution of
amino acid frequencies agrees with our result of the primordial
evolution of amino acid frequencies in principle. We believe that
the motivation of the amino acid gain and loss is the ``pressure''
by established pattern of the variation of amino acid frequencies
whose direction parallels with the direction of evolutionary flow.
The frequencies of P and A are just decrease rapidly, while the
frequencies of S and F increase. The magnitudes of gain for C and M
are considerable great, but the frequencies of them do not vary
obviously in the primordial evolution. This disagreement might be
explained by the suggestion that they are late-coming amino acids
\cite{late_aa1}\cite{late_aa2}.

\section*{Discussion}

``The starting point of a biological investigation will be
theoretical." \cite{G} Our work is a successful example to study
biology based on general principles. It must be emphasized that the
explanations in the above are systematical. We have explained those
phenomena in a unified theoretical framework. Why are there many
linear or quasi-linear relationships, such as those manifested in
Fig 1a, Fig. 2a, Fig. 2b, Fig. 4? The reason can be shown by the
linguistic model: there is only one adjustable variant, i.e. time
{\it t}, in the model which drives the evolution of all quantities.
Our work shows that the origin and evolution of life is determined
by a definite mechanism. Otherwise we can not imagine that all the
dots in the above figures align with lines or somewhat curves
obediently. If there existing a mechanism which governs the origin
and evolution of life, life is indeed a consequence of the evolution
of matter and is doomed to appear in the universe. If the linguistic
mechanism is universal, the life must be universal in the deep sky.

An important property of the model is that the parameters of amino
acid frequencies are constant, which indicates that the abundance of
amino acids in the surroundings had no time to change when forming
the differences of amino acid frequencies among species. Therefore
the variation of amino acid frequencies should develop during a
short time in the transition from the RNA world to the DNA world. We
conjecture that there being many life groups governed by different
possible linguistic systems at early time on this planet, but one
system prevailed over others spontaneously as symmetry breaking.
Thus selecting genetic code was a matter of chance. All the amino
acids in the surroundings were used up to generate proteins in terms
of the contemporary genetic code, while other linguistic systems
demised.

Crick's frozen-accident theory states that: ``...To account for it
being the same in all organisms one must assume that all life
evolved from a single organism (more strictly, from a single closely
interbreeding population).'' \cite{27} Our theory differs with
Crick's theory. We suggest that a linguistic system rather than a
species was selected by chance to produce the universal genetic
code. There is an analogy between our theory and the evolution of
natural language. Which language among the ``ecology'' of natural
languages become the international language was a matter of chance
in the history. Our theory can explain the exceptions to the
universal genetic code which are utilized in mitochondrial and in
principal genomes of certain species: they might be the antique of
other linguistic systems. This phenomenon is familiar in natural
languages. Many ancient languages have demised, but some ancient
grammar rules may keep in the contemporary language.

We suggest that there is an intrinsic relationship between the laws
of life and the laws of matter, which determines the proper scale of
us comparing with the scale of the earth. Information plays the
central role in biology as well as in physics \cite{28}\cite{29}.
There is an upper limit of information to store in a certain space
\cite{sciam}. Here we present a heuristic ideal experiment. Let a
biologist and a physicist drink together in a room, where enough
hard disks are stored so that the amount of the information in this
room is
just below the critical value of the upper limit. When
inspiring new ideas come to the brains of the two scientists
continuously, the amount of the information must be able to exceed
the critical value, while the mass in the room is conservative. The
volume of the room must inflate driven by the increasing
information. It infers that the information is equivalent to mass to
some extent. Therefore, the life fight for a decent space to live in
this universe for themselves. If a universe is too small, the
information created by all the life in this universe must drive it
to expand by some unknown mechanism. Maybe the accelerating
expansion of our universe is propelled partly by the activities of
the life. The genetic code bridges the world of matter and the world
of life. A general theory is expected to push down the wall between
physics and biology so that we can consider the phenomena of life
and matter in a unified theory.

\section*{Materials and methods}

\subsection*{}

{\bf Data collection.} The amino acid frequencies and average
protein lengths for 106 species (85 eubacteria, 12 archaebacteria, 7
eukaryotes and 2 viruses) are obtained based on the data in PEP on
URL: http://cubic.bioc.columbia.edu/pep. The GC contents are
obtained from Genome Properties system\cite{30}. These species are
representative for all the species on the earth in studying the
evolution of amino acid frequencies and genomic GC content.

\subsection*{}

{\bf Linguistic rules and parameters.} For three parts of the model:
(i) There are one initial tree and two auxiliary trees in tree
adjoining grammar. The leaf on the tree of adjoining grammar stands
for amino acid, which is determined by (ii). The variant t (the only
adjustable parameter) represents the probability for the replacement
in adjoining operation\cite{20}. (ii) D, P, F, Q, M, V, S and
termination code are at positions of inner node on the tree based on
genetic code multiplicity,  while G, A, E, L, T, R, I, N, H, K, C,
Y, W are at positions of leaf on the tree (Fig. 5). The 20
amino acid frequencies at present are input as constant parameters
in the model, which represent the amount of amino acids in
surroundings and determine the probabilities (also being constant)
of selecting amino acids at each fork of the tree. The amino acid to
join the protein sequences in (i) is selected randomly from root to
a leaf along the branches of the tree in first loop of the program
and is selected randomly from root or a stochastic inner node to a
leaf in the following loops. The probability of starting from an
inner node other than root in the following loops increases linearly
with t, which results in the evolution of amino acid frequencies in
simulation. (iii) The degenerate genetic codes are used from the
earliest to the latest chronologically\cite{22} for each amino acid
as t increases.

\begin{figure}
\centering{
\includegraphics[width=80mm, height=60mm]{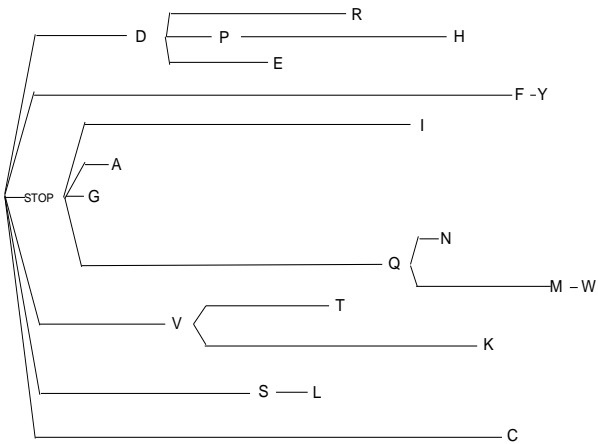}}
\label{fig5}
\caption{{\bf Grammar tree of genetic code multiplicity.} The 20 amino
acids are aligned chronologically from left to right. This picture
is faithfully base on the genetic code multiplicity in Ref.
\cite{21} and amino acid chronology in Ref. \cite{19}. The amino
acids are selected to join the protein sequence depended strictly on
this tree in the program for the model.}
\end{figure}

\subsection*{}

{\bf Simulation of sequence generation.} Starting from the initial
tree, protein sequence can be generated by adjoining operations.
There are two steps for the generation of protein in our model.
Mini- proteins with length about 10 aa are generated in the first
step; then the whole protein is obtained by connecting mini-proteins
in the second step (also based on tree adjoining grammar) when the
protein length increases proportionally. The amino acid frequencies
do not change in the second step. The simulations in Fig.
1, Fig. 2 and Fig. 4 are obtained by
calculating 50,000 generated mini-proteins to avoid stochastic error
for 30 species when {\it t} increases from 0.02 to 0.40 by equal
steps. Where the greater {\it t} is avoid because greater {\it t}
results in too long proteins in the program.

\subsection*{}

{\bf Protein length spectrum analysis.} Protein length spectrum is
obtained by counting the numbers of proteins with a certain length
in each of the 106 proteomes. Protein length spectra of three
domains are obtained by summing up all protein length spectra of the
species in each domain. Simulated protein length spectrum is
obtained when generating a certain amount of proteins.

PLSCC is defined as the inner product of a pair of normalized
protein length spectra (taking each protein length spectrum for a
vector), therefore it is just the cosine of the angle between the
two vectors of the spectra for the pair of species. Average PLSCC is
obtained by averaging the PLSCC's between a species and all the
other species (Fig. 3b). The $106\times106$ PLSCC matrix is
obtained by calculating PLSCC for each pair among 106 complete
proteomes (the species are sorted by average protein length). PLSCC
direction is defined as the direction along which average PLSCC's
for a group of closely related species (represented by overlapping
different shapes on stars in Fig. 4) decrease on the whole,
which reflects direction of evolution.

\section*{}

{\small We thank Shengli Zhang, Zhenwei Yao, Hefeng Wang, Qinye Yin
for discussions and Zhiwei Li and Philip Carter for assistances.}


\begin{thebibliography}{5}

\bibitem{3} Sueoka N (1961) {\it Proc. Natl. Acad. Sci. USA} {\bf 47}, 1141-1149.

\bibitem{4} Gu X, Hewett-Emmett D, Li WH (1998) {\it Genetica} {\bf 103}, 383-391.

\bibitem{8} Gilbert W (1986) {\it Nature} {\bf 319}, 618.

\bibitem{9} Joyce GF (2002) {\it Nature} {\bf 418}, 214-221.

\bibitem{10} Szathm\'{a}ry E (2003) {\it Nature Rev. Genetics} {\bf 4}, 995-1001.

\bibitem{11} Knight RD, Landweber LF (2000) {\it Cell} {\bf 101}, 569-572.

\bibitem{12} Wong JTF (1976) {\it Proc. Natl. Acad. Sci. USA} {\bf 73}, 2336-2340.

\bibitem{13} Wong JTF (1980) {\it Proc. Natl. Acad. Sci. USA} {\bf 77}, 1083-1086.

\bibitem{14} Szathm\'{a}ry E (1999) {\it Trends  Genet.} {\bf 15}, 223-229.

\bibitem{Jukes} Osawa S, Jukes TH, Watanabe K, Muto A (1992) {\it Microbiol. Rev.} {\bf 56}, 229-264.

\bibitem{27} Crick FHC (1968) {\it J. Mol. Biol.} {\bf 38}, 376-379.

\bibitem{periodic_table} Jungck JR (1978) {\it J. Mol. Evol.} {\bf 11}, 211-224.

\bibitem{5} Jordan IK et al (2005) {\it Nature} {\bf 433}, 633-638.

\bibitem{6} Hurst LD, Feil EJ, Rocha EPC (2006) {\it Nature} {\bf 442}, E11-E12.

\bibitem{1} Rost B (2002) {\it Curr. Opin. Stru. Biol.} {\bf 12}, 409-416.

\bibitem{2} Liu JF, Rost B (2001) {\it Protein Sci.} {\bf 10}, 1970-1979.

\bibitem{23} Muto A, Osawa S (1987) {\it Proc. Natl. Acad. Sci. USA} {\bf 84}, 166-169.

\bibitem{Sueoka 1962} Sueoka N (1962) {\it Proc. Natl. Acad. Sci. USA} {\bf 48}, 582-592.

\bibitem{15} Ji S (1997) {\it Biosynthesis} {\bf 44}, 17-39.

\bibitem{7} Gimona M (2006) {\it Nature Rev. Mol. Cell Boil.} {\bf 7}, 68-73.

\bibitem{16} Searls D (2002) {\it Nature} {\bf 420}, 211-217.

\bibitem{17} Editorial (2002) {\it Nature Struct. Biol.} {\bf 9}, 713.

\bibitem{18} Mantegna RN et al (1994) {\it Phys. Rev. Lett.} {\bf 73}, 3169-3172.

\bibitem{19} Trifonov EN (2004) {\it J. Biomol. Struct. Dyn.} {\bf 22}, 1-11.

\bibitem{D} Doolittle WF (1997) {\it Proc. Natl. Acad. Sci. USA} {\bf 94}, 12751-12753.

\bibitem{20} Joshi AK, Schabes Y (1997) in {\it Handbook of Formal
Languages}, eds Rozenberg G, Salomma A (Springer,Heidelberg),
pp.69-214.

\bibitem{21} Eduardo J, Hornos M, Hornos YMM (1993) {\it Phy. Rev. Lett.} {\bf 71}, 4401-4404.

\bibitem{22} Trifonov EN, Kirzhner A, Kirzhner VM, Berezovsky IN (2001) {\it J. Mol. Evol.} {\bf 53}, 394-401.

\bibitem{GC} Gorban A, Popova T, Zinovyev A (2005) {\it Physica A} {\bf 353}, 365-387.

\bibitem{24} Berman AL, Kolker E, Trifonov EN (1994) {\it Proc. Natl. Acad.
Sci. USA} {\bf 91}, 4044-4047.

\bibitem{25} Woese CR, Kandler O, Wheelis ML (1990) {\it Proc. Natl. Acad. Sci.
USA} {\bf 87}, 4576-4579.

\bibitem{26} Mayr E (1998) {\it Proc. Natl. Acad. Sci. USA} {\bf 95}, 9720-9723

\bibitem{late_aa1} Brooks DJ, Fresco JR, Lesk AM, Singh M (2002) {\it Mol. Biol. Evol.} {\bf 19},
1645-1655.

\bibitem{late_aa2} Brooks DJ, Fresco JR (2002) {\it Mol. Cell. Proteom.} {\bf 1},
125-131.

\bibitem{G} Gilbert W (1991) {\it Nature} {\bf 349}, 99.

\bibitem{28} Zurek WH ed (1990) {\it Complexity, Entropy and the Physics of Information} (Addison-
Wesley, Redwood City).

\bibitem{29} Susskind L, Lindesay J (2005) {\it An Introduction to Black Holes, Information, and the
String Theory Revolution: the Holographic Universe} (World
Scientific, Singapore).

\bibitem{sciam} Bekenstein JD (2003) {\it Scientific American} {\bf 289}, 58-65.

\bibitem{30} Haft DH, Selengut JD, Brinkac LM, Zafar N, White O (2005) {\it Bioinformatics} {\bf 21}, 293-306.


\end{thebibliography}
\end{document}